# Vacuum Technology for Superconducting Devices

*P. Chiggiato*[1]
CERN, Geneva, Switzerland

**Abstract**
The basic notions of vacuum technology for superconducting applications are presented, with an emphasis on mass and heat transport in free molecular regimes. The working principles and practical details of turbomolecular pumps and cryopumps are introduced. The specific case of the Large Hadron Collider's cryogenic vacuum system is briefly reviewed.

## 1  Introduction

Vacuum is necessary during the production of superconducting thin films for RF applications and for the thermal insulation of cryostats. On the other hand, vacuum systems take an advantage from the low temperatures necessary for superconducting devices. This chapter focuses on the principles and the main definitions of vacuum technologies; some insights about gas and heat transfer in a free molecular regime are given. Only turbomolecular pumping and cryopumping are described, since they are the most relevant for superconducting applications. Pressure measurement is not included because, in general, it is not considered a critical issue in such a domain. A comprehensive introduction to vacuum technology may be found in the books listed in the references.

## 2  Basic notions of vacuum technology

The thermodynamic properties of a rarefied gas are described by the ideal gas equation of state:

$$PV = N_{\text{moles}} RT \tag{1}$$

where $P$, $T$ and $V$ are the gas pressure, temperature, and volume, respectively, and $R$ is the ideal gas constant (8.314 J K$^{-1}$ mol$^{-1}$ in SI units); $N_{moles}$ is the number of gas moles. From statistical physics, Eq. (1) may be rewritten in terms of the total number of molecules $N$ in the gas:

$$PV = Nk_{\text{B}}T \tag{2}$$

where $k_\text{B}$ is the Boltzmann constant ($1.38 \times 10^{-23}$ J K$^{-1}$ in SI units). Pressure units are pascals (1 Pa = 1 N m$^{-2}$) or millibars (1 mbar = 100 Pa). Quantity of gas may be expressed as the number of molecules, number of moles or, by Eq. (1), pressure-volume units at a given temperature. The latter units are used in most cases in vacuum technology. The quantities of gas expressed in pressure-volume units are converted to the number of molecules by dividing them by $k_\text{B}T$ (1 Pa m$^3$ = $2.5 \times 10^{20}$, 1 mbar l = $2.5 \times 10^{19}$, both at 295 K).

In vacuum systems, pressures span several orders of magnitude (Table 1). *Degrees of vacuum* are defined by upper and lower pressure boundaries. Different degrees of vacuum are characterized by different pumping technologies, pressure measurement, materials, and surface treatments. In general, vacuum systems for particle accelerators operate in the high and ultra-high vacuum.

---

[1] paolo.chiggiato@cern.ch

Table 1: Degrees of vacuum and their pressure boundaries

|  | Pressure boundaries (mbar) | Pressure boundaries (Pa) |
|---|---|---|
| Low vacuum (LV) | 1000–1 | $10^5$–$10^2$ |
| Medium vacuum (MV) | 1–$10^{-3}$ | $10^2$–$10^{-1}$ |
| High vacuum (HV) | $10^{-3}$–$10^{-9}$ | $10^{-1}$–$10^{-7}$ |
| Ultra-high vacuum (UHV) | $10^{-9}$–$10^{-12}$ | $10^{-7}$–$10^{-10}$ |
| Extreme vacuum (XHV) | <$10^{-12}$ | <$10^{-10}$ |

The *mean speed of gas molecules* $\langle v \rangle$ is calculated by the Maxwell–Boltzmann distribution [1]:

$$\langle v \rangle = \sqrt{\frac{8 k_B T}{\pi \cdot m}} = \sqrt{\frac{8 RT}{\pi \cdot M}} \qquad (3)$$

where $m$ is the mass of the molecule and $M$ is the molar mass. The unit of both masses is kg in SI. Assuming that the density of molecules all over the volume is uniform, the *molecular impingent rate* φ onto a surface of unit area is calculated:

$$\phi = \frac{1}{4} n \langle v \rangle = \frac{1}{4} n \sqrt{\frac{8 k_B T}{\pi \cdot m}} \qquad (4)$$

where $n$ is the volume number density ($n = N\,V^{-1}$).

The shortest conceivable time to form a single layer of adsorbed molecules, i.e. when all molecules impinging on a surface adhere with it, is called the *monolayer formation time* $t_{ML}$. Assuming typically $10^{19}$ adsorption sites per m$^2$:

$$t_{ML}[s] = \frac{10^{19}}{\phi} \approx 93 \times 10^6 \frac{\sqrt{mT}}{P}. \qquad (5)$$

For nitrogen ($m = 4.6 \times 10^{-26}$ kg) at room temperature and pressures of the order of $10^{-4}$ Pa, $t_{ML}$ is of the order of 1 s.

In any physically limited vacuum system, molecules collide between each other and with the walls of the vacuum envelope/container. When the gas density is so low that molecular collisions are improbable, there is a drastic change in the transport phenomena. This important feature is described by the Knudsen number $K_n$, namely the ratio between the average distance between two consecutive points of molecular collision $\bar{\lambda}$ [1], i.e. the *molecular mean free path*, and a characteristic dimension $D$ of a vacuum system, for example the diameter in cylindrical beam pipes:

$$K_n = \frac{\bar{\lambda}}{D}. \qquad (6)$$

As reported in Table 2, the values of $K_n$ delimit three different gas flows: free molecular, viscous, and transitional.

**Table 2:** Gas dynamic regimes defined by the Knudsen number

| $K_n$ range | Regime | Description |
|---|---|---|
| $K_n > 0.5$ | Free molecular flow | Molecule–wall collisions dominate |
| $K_n < 0.01$ | Continuous (viscous) flow | Gas dynamics dominated by intermolecular collisions |
| $0.5 < K_n < 0.01$ | Transitional flow | Transition between molecular and viscous flow |

In typical beam pipes ($D \sim 10$ cm), the transition to molecular regime is obtained for pressures lower than about $10^{-1}$ Pa. Vacuum systems of accelerators operate in the free molecular regime, except for short transfer lines and ion sources. The free molecular flow regime characterizes and determines the pumping and pressure-reading mechanisms. Since there is no interaction between molecules, pumps and instruments must act on single molecules.

In stationary conditions, the number densities ($n_1$ and $n_2$) and pressures ($P_1$ and $P_2$) in two connected volumes at different temperatures ($T_1$ and $T_2$) are correlated by means of Eq. (7):

$$\frac{n_1}{n_2} = \sqrt{\frac{T_2}{T_1}}; \quad \frac{P_1}{P_2} = \sqrt{\frac{T_1}{T_2}}. \tag{7}$$

In general, it can be written that:

$$\frac{P}{\sqrt{T}} = \text{constant}; \quad n\sqrt{T} = \text{constant}. \tag{8}$$

Therefore, in equilibrium, in a vessel at lower temperature there is a higher density and a lower pressure. With reference to Fig. 1, Eq. (7) can be obtained by equating the impingement rate (see Eq. (4)) from volume 1 to volume 2 and vice versa. This phenomenon is usually referred to as *thermal transpiration*.

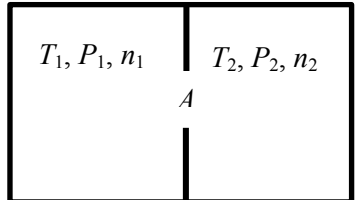

**Fig. 1:** Schematic drawing of two volumes communicating through a thin, small wall slot

## 2.1 Gas transport in free molecular flow

### 2.1.1 Gas conductance

In the free molecular regime and isothermal conditions, the net gas flow $Q$ between two points of a vacuum system is proportional to the pressure difference ($P_1 - P_2$) at the same points:

$$Q = C(P_1 - P_2) \tag{9}$$

where $C$ is called the gas conductance of the vacuum system between the two points. In the free molecular regime, the conductance does not depend on pressure. It depends only on the mean molecular speed and the vacuum system geometry. If the gas flow units are expressed in terms of pressure–volume (e.g. mbar l s$^{-1}$ or Pa m$^3$ s$^{-1}$), the conductance is reported as volume per unit time (i.e. l s$^{-1}$ or m$^3$ s$^{-1}$) (litre is represented by the letter l).

The conductance of a wall slot of area $A$ between two vessels is given by Eq. (10). It can be obtained by calculating the difference between the impingement rates for volumes 1 and 2 in isothermal conditions, with reference to Fig. 1.

$$C = \frac{1}{4}A\langle v \rangle \propto A\sqrt{\frac{T}{m}} \tag{10}$$

Therefore, the conductance of the wall slot is inversely proportional to the square root of the molecular mass. For equal pressure drop, the gas flow of $H_2$ is the highest. The conductances $C'$ of a unit-surface-area wall slot for different gas molecules are listed in Table 3.

**Table 3:** Wall slot conductances $C'$ of 1 cm$^2$ surface area, for common gas species, at room temperature

| Gas | $H_2$ | He | $CH_4$ | $H_2O$ | $N_2$ | Ar |
|---|---|---|---|---|---|---|
| $C'$ at 293 K (l s$^{-1}$ cm$^{-2}$) | 44 | 31.1 | 15.5 | 14.7 | 11.75 | 9.85 |

For complex geometries, the transmission probability $\tau$ is introduced. The conductance of any duct connecting two vessels at the same temperature is given by Eq. (11).

$$C = C'A_1 \tau_{1 \to 2} \tag{11}$$

where $C'A_1$ is the conductance of the duct aperture seen as a wall slot in the first vessel, and $\tau_{1 \to 2}$ is the probability that a molecule once entered into the duct from the first vessel is transmitted to the other vessel.

The transmission probabilities depend only on vacuum component geometry. Approximate formulas are reported in many vacuum technology textbooks, for example in Refs. [2–5]. For long tubes ($D \gg L$, where $D$ is diameter and $L$ is length) the transmission probability and the resulting conductance are:

$$\tau \approx \frac{4}{3}\frac{D}{L} \tag{12}$$

$$C \approx 12.3 \frac{D^3}{L}. \tag{13}$$

In Eq. (13) $D$ and $L$ are expressed in cm, $C$ in l s$^{-1}$. Conductances of more complicated components are calculated by test-particle Monte Carlo methods (TPMC). The reference TPMC software at CERN is MolFlow+ [6].

Vacuum components are installed either in series, i.e. traversed by the same gas flow, or in parallel, i.e. equal pressures at the extremities. In the former case, the inverse of the total conductance $C_{TOT}$ is given by the sum of the inverse of the single conductances $C_i$:

$$\frac{1}{C_{TOT}} = \sum_{1}^{N}\frac{1}{C_i}. \tag{14}$$

In the latter case, the total conductance is the sum of the single conductances:

$$C_{TOT} = \sum_{1}^{N} C_i. \tag{15}$$

### 2.1.2 Pumping speed

In vacuum technology, a pump is any 'object' that removes gas molecules from the gas phase. A vacuum pump is characterized by its pumping speed $S$, which is defined as the ratio between the pumped gas flow $Q_P$ (pump throughput) and the pump inlet pressure $P$:

$$S = \frac{Q_P}{P}. \qquad (16)$$

The pump throughput can be written as the gas flow through the cross-section of the pump inlet (surface area $A_P$) multiplied by the capture probability $\sigma$, i.e. the probability for a molecule that enters the pump to be definitely removed and never more to reappear in the gas phase of the vacuum system (see Eq. (17)). In the literature, $\sigma$ is also called the Ho coefficient:

$$Q_P = \phi A_P \sigma = \frac{1}{4} A_P n \langle v \rangle \sigma. \qquad (17)$$

Considering Eqs. (2) and (10), it turns out that:

$$Q_P = A_P C' \frac{P}{k_B T} \sigma. \qquad (18)$$

From the definition of the pumping speed and converting the throughput into pressure–volume units:

$$S = A_P C' \sigma. \qquad (19)$$

Therefore, the pumping speed is equal to the conductance of the pump inlet cross-section multiplied by the capture probability. The maximum theoretical pumping speed of any pump is obtained for $\sigma = 1$ and it is equal to the conductance of the pump inlet cross-section. Table 4 reports some values of the maximum pumping speed of lump pumps for typical diameters of the pump inlet. Because $S$ depends on $C'$, and so on the inverse of the square root of the molecular mass, the maximum theoretical pumping speed is that for $H_2$.

The pumping speed given by the suppliers of vacuum pumps is called the *nominal pumping speed*; it refers to the pump inlet. The *effective pumping speed* $S_{eff}$ is that acting directly in the vacuum vessel of interest. The effective pumping speed is lower than the nominal pumping speed owing to gas flow restrictions interposed between the pump and the vessel.

**Table 4:** Maximum pumping speeds in l s$^{-1}$ for different diameters of circular pump inlets

| ID (mm) | H$_2$ (l s$^{-1}$) | N$_2$ (l s$^{-1}$) | Ar (l s$^{-1}$) |
|---|---|---|---|
| 36 | 448 | 120 | 100 |
| 63 | 1371 | 367 | 307 |
| 100 | 3456 | 924 | 773 |
| 150 | 7775 | 2079 | 1739 |

The effective pumping speed is calculated equating the net gas flow from the vessel and that removed by the pump. Taking into account Eqs. (9) and (16), with reference to Fig. 2, one obtains:

$$Q = C_1(P_1 - P_2) = SP_2 = S_{eff} P_1 \qquad (20)$$

and so:

$$\frac{1}{S_{eff}} = \frac{1}{S} + \frac{1}{C}. \qquad (21)$$

As a result, for $C \ll S$, one finds $S_{\text{eff}} \approx C$. In other words, the effective pumping speed does not depend on the installed pump if the conductance of the interposed connection is very low. This conclusion is of primary importance in the design of efficient vacuum systems.

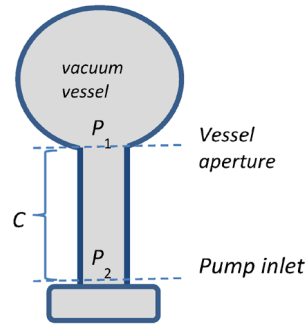

**Fig. 2:** Schematic drawing of a gas flow restriction of conductance $C$ interposed between a pump of pumping speed $S$ and a vacuum vessel.

### 2.1.3 Outgassing

Several gas sources contribute to the total gas load in a vacuum system. In addition to intentional gas injections and air leaks, outgassing of materials plays a crucial role in the global gas balance. Materials release gas molecules that are both adsorbed onto their surfaces and dissolved in their bulk [7]. The outgassing is distributed uniformly all over the vacuum-exposed surfaces; its rate is in general reported for unit surface area ($q$).

The outgassing rate is reduced by dedicated surface treatments. For metals, in general, several surface treatments are available, each aiming at removing a specific set of contaminants. Gross contamination and the sorption layer (hydrocarbons, Cl compounds, silicone greases, etc.) are removed by solvent and detergent cleaning. Thick oxide layers (1–10 nm thick) are removed by chemical pickling. The damaged metal skin – full of dislocations, voids, and impurities – is removed by acid etching or electropolishing.

For all materials, after thorough surface treatment, water vapour dominates the outgassing process. Its outgassing rate is inversely proportional to the pumping time ($t$) for smooth metals. The following empirical relationship is in general applied:

$$q_{\text{H}_2\text{O}} \approx \frac{3 \times 10^{-9}}{t[\text{h}]} \left[\frac{\text{mbar l}}{\text{s cm}^2}\right]. \tag{22}$$

For organic materials, in particular polymers, in the first phase of pumping, the water vapour outgassing rate is inversely proportional to the square root of the pumping time $q_{\text{H}_2\text{O}} \propto 1/\sqrt{t}$, the absolute value being strongly dependent on the nature and history of the material. For example, the Multi-Layer Insulation (MLI) used for the LHC insulation vacuum is made of polymer foils. There are several such layers amounting to 250 m² of MLI per metre of the LHC. When a typical insulation vacuum sector is exposed to ambient air for several weeks, it takes 200 hours of pumping with an effective pumping speed of 100 l s$^{-1}$ to attain a pressure of about $5 \times 10^{-3}$ mbar at room temperature; from such data an outgassing rate of $10^{-9}$ mbar l s$^{-1}$ cm$^{-2}$ per layer of MLI is obtained. Such a pressure would have been achieved, after the same pumping time, with a stainless steel sheet as wide as two soccer fields per metre of LHC sector.

The water vapour outgassing is reduced by *in situ* heating in vacuum (bake-out). Such a thermal treatment is very effective for metals if it is carried out for at least 12 h at temperatures higher than 120°C. After bake-out, H$_2$ becomes the main outgassed species for metals; the outgassing rate can be regarded as constant at room temperature. It can be reduced by heat treatment at higher temperatures either *in situ,* or *ex situ* in a vacuum furnace [7].

As shown in Table 5, the value of outgassing rate spans several orders of magnitude. The right choice of materials and treatments is essential in the design of vacuum systems to limit the gas charge. The outgassing features should be taken into account at the very beginning of the design process when materials are selected.

**Table 5:** Values of outgassing rates for selected materials used in vacuum technology. The reported heat treatments are carried out both *in situ* and in vacuum.

| Material | $q$ (mbar l s$^{-1}$ cm$^{-2}$) | Main gas species |
|---|---|---|
| Neoprene, not baked, after 10 h of pumping [8] | Order of $10^{-5}$ | $H_2O$ |
| Viton, not baked, after 10 h of pumping [8] | Order of $10^{-7}$ | $H_2O$ |
| Austenitic stainless steel, not baked, after 10 h of pumping | $3 \times 10^{-10}$ | $H_2O$ |
| Austenitic stainless steel, baked at 150°C for 24 h | $3 \times 10^{-12}$ | $H_2$ |
| OFS copper, baked at 200°C for 24 h | Order of $10^{-14}$ | $H_2$ |

In particle accelerators, gas molecules can also be desorbed by electron, photon, and ion collisions induced by the beam. In modern high-energy accelerators, such a gas source is generally preponderant. Significant reduction of desorption yields is obtained by gradually accumulating high doses of particle collisions. Another very efficient method is coating the inner surface of the beam pipe with non-evaporable getter thin films that, after *in situ* activation, achieve a very clean and pumping surface.

### 2.1.4 Pressure profiles

The calculation of the pressure profile along vacuum systems is an essential task of vacuum experts and should be tackled at the design phase. In general, the contributions to the total pressure of localized and distributed gas sources are considered separately and finally added. This is possible because in most cases the equations that describe pressure profiles are linear. This may not be true if the pumping speed is pressure-dependent.

In the case of a localized gas source, the pressure in a vacuum vessel is obtained by taking into account Eq. (16) and the intrinsic pressure limitation $P_0$ of the installed pumping system:

$$P = \frac{Q}{S} + P_0. \tag{23}$$

The pumping speed $S$ is given either by the supplier or preliminarily measured; $P_0$ is the pressure attained in the system without any gas load. When a restriction of conductance $C$ is interposed between the pump and the vessel, the effective pumping speed $S_{eff}$ is considered and Eq. (23) becomes:

$$P = \frac{Q}{S_{eff}} + P_0 = \frac{Q(C+S)}{CS} + P_0. \tag{24}$$

When many vessels are interconnected, the flow balance is written in each vessel (node analysis). This analysis leads to a system of linear equations, from which the pressure values in each vessel are calculated. As an example, with reference to Fig. 3, in the first vessel, the injected gas flow ($Q$) is either pumped ($P_1S_1$) or transmitted to the second vessel ($C_1(P_1 - P_2)$). This latter flow is pumped in the second vessel or transmitted to the third vessel, and so on. Thus:

$$\begin{aligned} Q &= P_1 S_1 + C_1(P_1 - P_2), \\ C_1(P_1 - P_2) &= C_2(P_2 - P_3) + P_2 S_2, \\ C_2(P_2 - P_3) &= C_3(P_3 - P_4) + P_3 S_3, \\ C_3(P_3 - P_4) &= P_4 S_4. \end{aligned} \qquad (25)$$

When a second localized gas flow is settled, the node analysis is repeated. The contributions of each localized gas flow to the pressure values are then added. If the cross-section of the interconnection ducts is constant, it can be shown that the pressure varies linearly between two connected vessels.

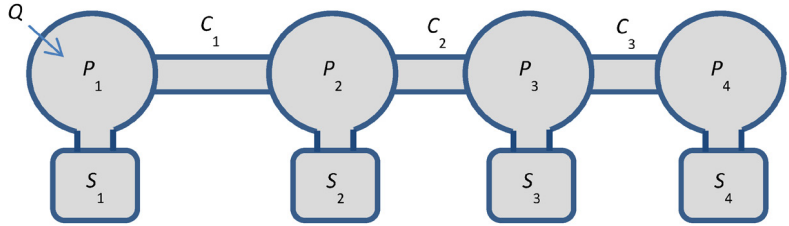

**Fig. 3:** Schematic drawing of four interconnected vacuum vessels. In each vessel, the gas flow balance is written (node analysis).

In case of distributed outgassing, a local gas balance is written. The resulting equations are typical of transport phenomena:

$$\begin{aligned} q(x) &= -\bar{C}(x) \frac{\partial P(x,t)}{\partial x}, \\ \bar{C}(x) \frac{\partial^2 P(x,t)}{\partial x^2} &= \frac{\partial P(x,t)}{\partial t}. \end{aligned} \qquad (26)$$

where $q(x)$ is the gas flow at abscissa $x$. $\bar{C}(x)$ is the conductance of the unit length of a vacuum vessel having a constant cross-section as that at abscissa $x$; $\bar{C}(x)$ is constant and equal to $CL$ if the vessel has a constant cross-section, total conductance $C$, and length $L$. In a steady state, the second space derivative of the pressure is zero; consequently, the pressure profiles are parabolic. A typical example is a uniform beam pipe, with distributed outgassing, that is pumped at both extremities by lump pumps. The pressure reaches a maximum $P_{max}$ in the middle of the beam pipe and minimum $P_{min}$ at the pumping positions. The difference $\Delta P$ between the two extremes is given by:

$$\Delta P = P_{max} - P_{min} = \frac{Q}{8C} \qquad (27)$$

where $Q$ is the total outgassing rate of the beam pipe. Nowadays, pressure profiles are also easily obtained by TPMC simulation.

## 2.2 Heat transport by gas

Gas molecules participate in the global heat transfer process transferring energy from the warmer walls to the colder ones. The contribution of gas convection is erased at relatively high pressures; for this reason it is not considered here.

The energy that a gas molecule can convey depends on how the energy is stored in the molecule, i.e. on the molecular movements that are active at a certain temperature. Translational movements in the three dimensions are common to all molecules. Molecules formed with more than one atom can also rotate around the centre of mass. Vibrational modes are in general inactive at cryogenic temperature. Kinetic theory shows that the thermal energy is equally shared among each active molecular movement: each active degree of freedom carries the same average energy, which is equal to ½$k_B T$.

The specific heat capacity at constant volume ($c_v$) of a *single molecule* with $f$ active degrees of freedom is:

$$c_v = f \frac{k_B}{2m}. \qquad (28)$$

The *molar* specific heat capacity at constant volume ($c_{Mv}$) is obtained from Eq. (28) multiplied by the mass of one mole, i.e. $N_A m$:

$$c_{Mv} = f \frac{N_A k_B}{2} = f \frac{R}{2}. \qquad (29)$$

For an ideal gas, specific heat capacities at constant pressure ($c_p$ and $c_{Mp}$) can be calculated [9]:

$$c_p = c_v + \frac{k_B}{m}, \qquad c_{Mp} = c_{Mv} + R. \qquad (30)$$

The specific heat capacity ratio is known as the isentropic expansion factor $\gamma$:

$$\gamma = \frac{c_p}{c_v} = \frac{c_{Mp}}{c_{Mv}} = 1 + \frac{2}{f}. \qquad (31)$$

### 2.2.1 Gas thermal conduction in viscous regime

In a viscous regime, the conduction heat flow is given by the Fourier equation:

$$q_{th} = -k_{th} \nabla T. \qquad (32)$$

The gas thermal conductivity $k_{th}$ is given by Eq. (33):

$$k_{th} = \frac{1}{4}(9\gamma - 5) \cdot \eta \cdot c_v \qquad (33)$$

where $\eta$ is the gas viscosity:

$$\eta = \frac{5}{16} \frac{1}{\delta^2} \sqrt{\frac{m k_B T}{\pi}} \qquad (34)$$

and $\delta$ is the molecular diameter. Table 6 reports selected values of $\delta$, $m$, and $f$ for simple molecules.

**Table 6:** Mass, diameter, and active degree of freedom at room temperature (translational and rotational) for typical gas molecules found in HV and UHV.

|  | $m$ (kg) | $\delta$ (pm) | $f$ |
|---|---|---|---|
| $N_2$ | $4.6 \times 10^{-26}$ | 370 | 5 |
| $H_2$ | $3.3 \times 10^{-27}$ | 293 | 5 |
| $H_2O$ | $3.0 \times 10^{-26}$ | 275 | 6 |
| CO | $4.6 \times 10^{-26}$ | 376 | 5 |
| $CH_4$ | $2.7 \times 10^{-26}$ | 382 | 6 |

The heat flow between two parallel plates, at temperatures $T_1$ and $T_2$ and distance $L$ may be approximately obtained by Eq. (35):

$$|q_{th}| \approx k_{th} \frac{T_1 - T_2}{L} \qquad (35)$$

where the gas conductivity is calculated for the temperature of the warmer plate. For example, the heat flow between two plates at 300 K and 80 K at a distance of 10 cm is about 55 Wm$^{-2}$ when the residual gas is nitrogen; the thermal conductivity is 25.3 mW K$^{-1}$m$^{-1}$.

The conduction heat flow in viscous regimes does not depend on the gas pressure: lower pressure means less energy carriers but a longer mean free path; $q_{th}$ is inversely proportional to the distance between the walls of the vessel.

### 2.2.2 Gas thermal conduction in molecular regime

In the molecular regime, gas molecules interact only with the walls of the vessel. The molecules carry thermal energy directly from one wall to another. The fraction of energy transferred after collision depends on the temperature and on the nature of the molecules and surfaces. The transfer process is quantitatively described in terms of a coefficient of thermal accommodation $\alpha$, which is a measure of the efficiency of energy exchange between a gas flow and a solid surface:

$$\alpha_1 = \frac{E_2' - E_1'}{E_2 - E_1} \text{ (for surface 1)},$$

$$\alpha_2 = \frac{E_1' - E_2'}{E_1 - E_2} \text{ (for surface 2)}. \qquad (36)$$

In Eq. (36), $E_2'$ and $E_1'$ are the thermal energies leaving, each second, the unit area of surface 2 and surface 1, respectively; $E_1$ and $E_2$ are the values that $E_1'$ and $E_2'$ would have if thermal equilibrium were achieved in all collisions at surfaces 1 and 2, respectively. In other words, the coefficient of thermal accommodation is the average fraction of the maximum possible energy exchange at the walls.

The equation for the heat flow in molecular regime was given by Kennard [2]; the subscript 2 refers to the warmer surface:

$$q_{th} = \alpha \frac{\gamma + 1}{\gamma - 1} \sqrt{\frac{R}{8 M \pi}} \frac{T_2 - T_1}{\sqrt{T}} P. \qquad (37)$$

Therefore, the heat flow is proportional to the pressure and does not depend on the distance between the two walls.

Here, $\alpha$ is the overall accommodation coefficient:

$$\alpha = \frac{\alpha_1 \alpha_2}{\alpha_2 + \alpha_1 (1 - \alpha_2) \frac{A_1}{A_2}}. \qquad (38)$$

Eq. (37) is valid for concentric spheres, coaxial cylinders, or parallel planes. In the case of two parallel slabs, of the same area, made of the same material, Eq. (38) simplifies to:

$$\alpha = \frac{\alpha_1}{2 + \alpha_1}. \qquad (39)$$

Thermal accommodation coefficients for most materials with technical surface finishing are about 0.9, 0.8, and 0.4 for argon, nitrogen, and helium, respectively (Table 7) [2, 10, 11, 12]. If an adsorbed or condensed layer of gas is formed, $\alpha$ tends to 1. The lower values of $\alpha$ for the lightest molecules counteract the mass dependence of Eq. (37) so that the same magnitude of heat flow is expected for He and $H_2$ as for $N_2$ on technical surfaces. This is not the case in a viscous regime where the thermal conductivity of $H_2$ is about one order of magnitude higher than that of nitrogen.

**Table 7:** Selected values of the thermal accommodation coefficient for platinum and stainless steel AISI-304 at room temperature [2, 10, 11, 12].

| Gas | $\alpha$ | | |
| --- | --- | --- | --- |
| | Pt technical surface at room temperature | AISI 304 technical surface | Pt atomically clean surface |
| $N_2$ | 0.77 | 0.8 | |
| $H_2$ | 0.29 | | 0.15 |
| Ar | 0.86 | 0.87 | 0.55 |
| CO | 0.78 | | |
| He | 0.38 | 0.36 | 0.03 |

The temperature $T$ in the denominator of Eq. (37) is an effective temperature of the non-equilibrium gas [13]. If the pressure $P$ is measured between the two walls, $T$ is given by the following equation:

$$\frac{1}{\sqrt{T}} = \left(\frac{A_1}{A_1 + A_2}\right)\frac{1}{\sqrt{T_1'}} + \left(\frac{A_2}{A_1 + A_2}\right)\frac{1}{\sqrt{T_2'}} \tag{40}$$

where $T_1'$ and $T_2'$ are the effective temperatures of the gas molecules after collision with surface 1 and 2, respectively. They can be expressed in terms of the real temperatures of the two walls, i.e. $T_1$ and $T_2$.

$$T_1' = \frac{\alpha_1 T_1\left[(1-\alpha_2)\frac{A_1}{A_2} + \alpha_2\right] + \alpha_2(1-\alpha_1)T_2}{\alpha_2 + \alpha_1(1-\alpha_2)\frac{A_1}{A_2}},$$

$$T_2' = \frac{\alpha_1 T_1(1-\alpha_2)\frac{A_1}{A_2} + \alpha_2 T_2}{\alpha_2 + \alpha_1(1-\alpha_2)\frac{A_1}{A_2}}. \tag{41}$$

Alternatively, if the pressure $P$ is measured by an external gauge, $T$ is the temperature at the gauge position, generally room temperature. In fact, because the ratio $P/\sqrt{T}$ is constant throughout the vacuum system, the values of $P$ and $T$ must be chosen at the same location.

As an example, two plates at 300 K and 80 K at a distance of 10 cm are considered. As in Section 2.2.1, the residual gas is $N_2$. The heat flow in the molecular regime is given by the approximated equation:

$$q_{th} \approx 265 \cdot P \,(\text{W m}^{-2}). \tag{42}$$

The heat radiation contribution calculated by the Stefan-Boltzmann equation gives a value of 8.9 W m$^{-2}$; to be realistic, values of thermal emissivity equal to 0.4 (stainless steel) and 0.02 (clean copper) are chosen for the surfaces at 300 K and 80 K, respectively. As a result, the gas conduction contributes to the same extent as radiation when the pressure is in the $10^{-2}$ Pa range. To reduce the gas heat transport to less than 10 times the radiation value, pressure in the $10^{-3}$ Pa range are needed. Figure 9 shows the different heat transport contributions, including in the viscous regime.

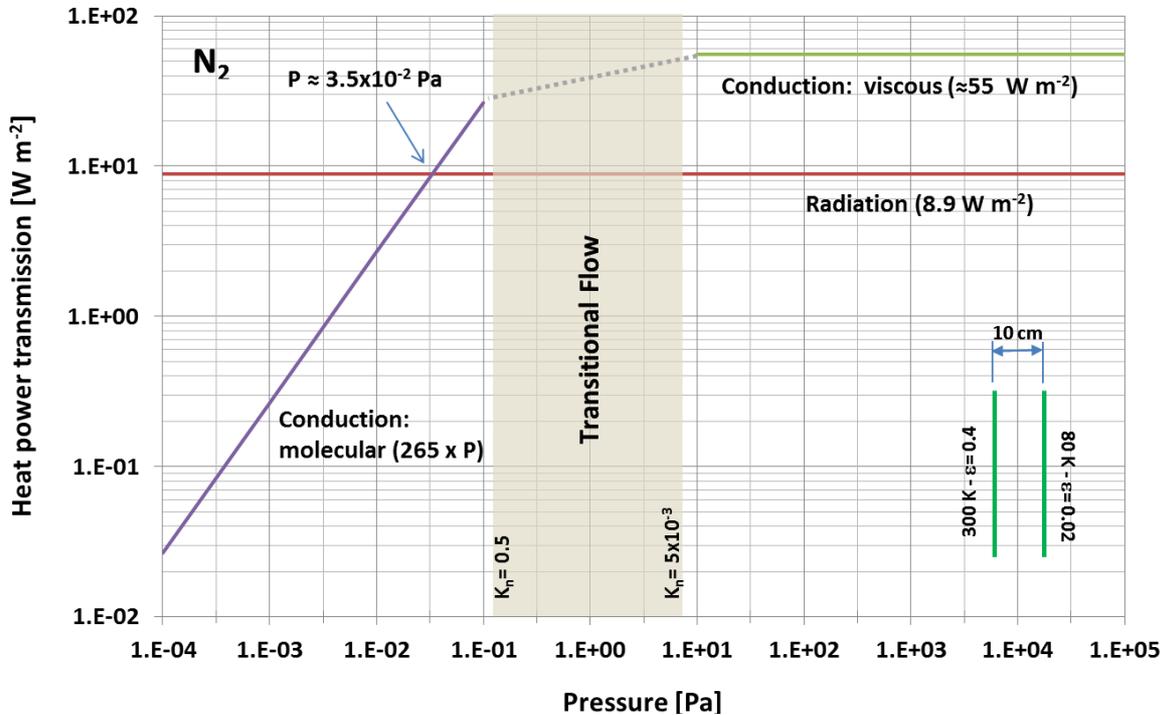

**Fig. 9:** Heat power transmission between two parallel plates at a mutual distance of 10 cm. The temperatures of the plates are 300 K and 80 K. The plates are made of stainless steel (thermal emissivity 0.4) and clean copper (thermal emissivity 0.02), for the warmer and colder surfaces, respectively. The residual gas is nitrogen.

## 3    Gas pumping in molecular regimes

Vacuum pumps in the molecular regime are classified into two families: momentum transfer pumps [2] and capture pumps [14]. Both act on each molecule singularly since no momentum and energy transfer is possible between molecules in this pressure range. In the first family, molecules receive a momentum component pointing towards the pump outlet (foreline) where the gas is compressed and evacuated by pumps working in the viscous regime (e.g. rotary vane, diaphragm, and scroll pumps). In this set of pumps, the turbomolecular ones are the most used for superconducting applications. The second family removes gas molecules by fixing them on a surface exposed to the vacuum. The molecular adsorption is based on chemical bonding (getter pumps) or physisorption at cryogenic temperatures. The latter may be a by-product of cooling at temperatures lower than the superconducting transition; it is the dominant pumping principle in beampipes embedded in superconducting magnets. Only turbomolecular pumps and cryopumping are considered in this chapter.

### 3.1    Turbomolecular pumps

In turbomolecular pumps (TMP), gas molecules are steered by rapidly rotating blades. The rotating blades are tilted with respect to the rotational axis. As a result, oblique channels are formed between successive blades. The rotor is inserted between two static surfaces defining two interspaces (see

Fig. 10). When a molecule comes from interspace-1, the angular distribution of the molecular velocity seen from the blades is preferentially oriented towards the blades' channels. Conversely, for those coming from interspace-2, the velocity vectors point preferentially towards the blades' wall, therefore increasing backscattering. A gas flow is consequently generated from interspace-1 to interspace-2. This mechanism works only if the angular distribution of the molecular speed as seen from the blade is significantly deformed, i.e. only if the peripheral speed of the blades is at least of the order of the mean molecular speeds (hundreds of metres per second).

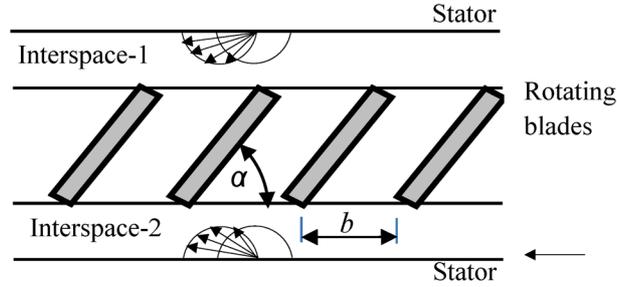

**Fig. 10:** A single rotor–stator stage in a turbomolecular pump, showing the molecular speed distributions as seen from the moving blades.

In a real TMP, the gas is compressed by several series of rotating blades. Each series of rotating blades is followed by a series of static blades. Molecules transmitted through the rotating blades' channels hit the static blades; as a result, the angular distribution of velocity is randomized and the molecules are ready for the next compression stage. The momentum transfer is effective only if the molecules do not experience intermolecular collisions after hitting the blades; this is equivalent to saying that the mean free path has to be larger than the blade distance. As a result, this type of pump works at full pumping speed only in molecular regimes ($P < 10^{-3}$ mbar).

The most important characteristics of a molecular pump are the pumping speed $S$ and the maximum compression pressure ratio that can be achieved between the pump inlet $P_{IN}$ and outlet $P_{OUT}$:

$$K_0 = \left(\frac{P_{OUT}}{P_{IN}}\right). \tag{43}$$

With reference to Fig. 10, it can be shown [2] that:
$$S \propto b \cdot u \cdot \sin\alpha \cos\alpha \, , \tag{44}$$

$$K_0 \propto \exp\left(\frac{u}{\langle v \rangle \cdot b \cdot \sin\alpha}\right) \propto \exp\left(\frac{u\sqrt{m}}{b \cdot \sin\alpha}\right). \tag{45}$$

To maximise the pumping speed, the distance between the blades is larger for the first series of blades and their inclination is 45°. The maximum compression ratio is obtained by modifying the geometry of the following series of blades; this is achieved by reducing both the distance between the blades and their inclination (see Fig. 11). In addition, the tighter blades compensate for the increased gas density and thus for the resulting lower molecular mean free path.

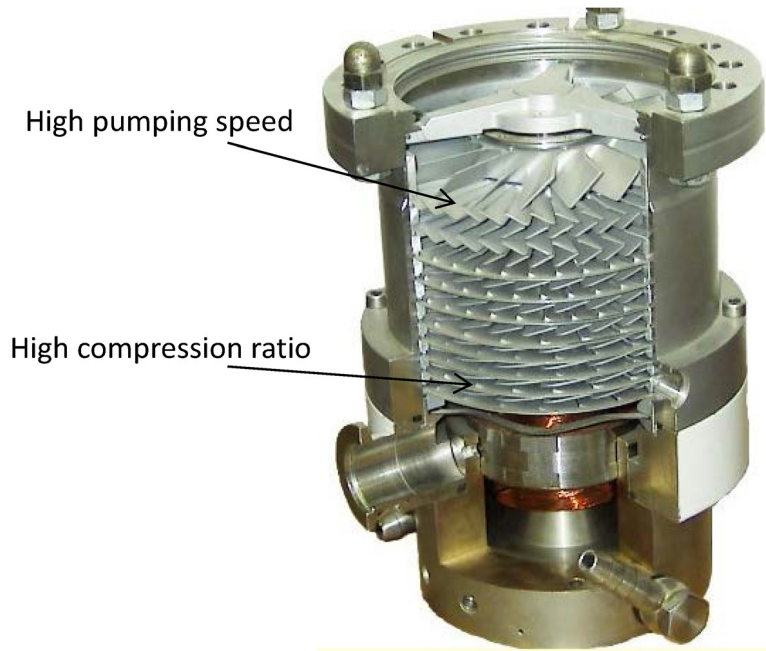

**Fig. 11:** Cut-out view of a turbomolecular pump. Courtesy of Wikipedia
http://en.wikipedia.org/wiki/Turbomolecular_pump).

The pumping speeds of commercial turbomolecular pumps (Fig. 12) are constant in the free molecular regime ($P < 10^{-3}$ mbar) and range between 10–3000 l s$^{-1}$ depending on pump inlet diameter and mechanical design. As indicated by Eq. (44), pumping speed should not be affected by the mass of the gas molecules. Indeed, in real pumps the gas dependence of the pumping speed is limited. In terms of pumping speed, turbomolecular pumps are not selective.

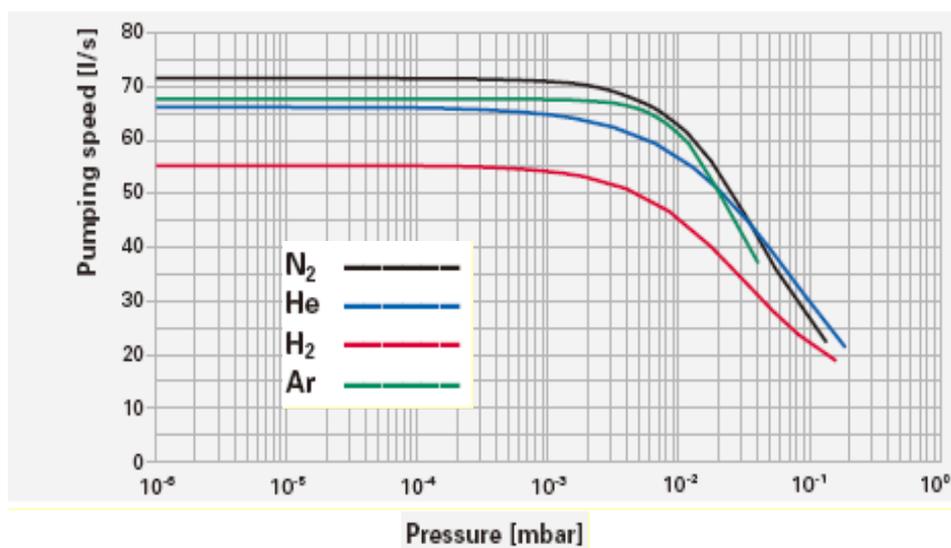

**Fig. 12:** Pumping speeds of a turbomolecular pump equipped with a DN63CF flange. For a TMP, the pumping speeds are constant in the free molecular regime and the nature of the pumped gas has a limited effect, if compared with other type of pumps. (Courtesy of Pfeiffer Vacuum.)

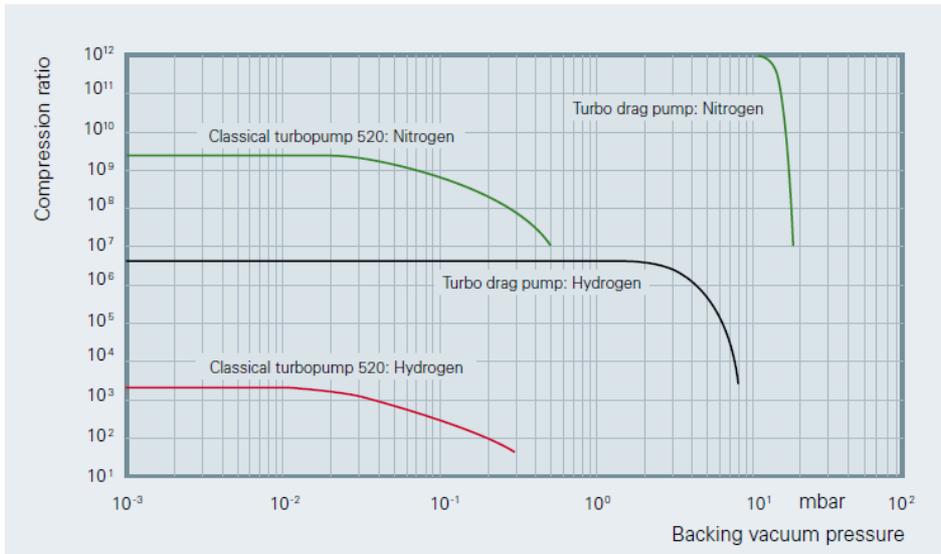

**Fig. 13:** Maximum compression ratio of two different turbomolecular pumps for $N_2$ and $H_2$. (Courtesy of Pfeiffer Vacuum.)

As expected, the maximum compression ratio is the lowest for $H_2$; in classical designs it is about $10^3$ (see Fig. 13). Currently, values up to $10^6$ are achieved by integrating a molecular drag pump (Gaede or Holweck drag stage) to the rear of the blade sets [2]. The TMP ultimate pressure is of the order of $10^{-10}$–$10^{-11}$ mbar for baked and all-metal vacuum systems. Lower pressures may be obtained if another type of pump removes the TMP $H_2$ back-streaming.

At the pump outlet, the compressed gas is evacuated by mechanical pumps (often called backing pumps) operating in the viscous regime. The turbomolecular pump and its backing pump are in general assembled in a single unit that includes power supplies, controls, and instrumentation.

The TMP can evacuate high gas flow at relatively high pressure without selectivity and memory effects. For example, a DN100 TMP can withstand a continuous gas flow up to $2 \times 10^{-1}$ mbar l s$^{-1}$ without damage.

The use of dry pumps for TMP backing has surmounted the risk of oil back-streaming from rotary vane pumps. The complete removal of all lubricated mechanical bearings has been obtained by magnetic rotor suspension.

The main drawback of a TMP is related to possible mechanical failures leading to definitive damage to the high-speed rotor. In addition, in case of unwanted rotor deceleration caused by a power cut or the rotor seizing, the vacuum system has to be protected by safety valves and dedicated pressure sensors against air back-stream. TMPs have limited application in radioactive environments due to possible damage to their electronics and power supplies. Some radiation-resistant TMPs are now available; for a significant cost, the whole of the electronics are moved beyond the radiation shielding by means of long cables. The application of TMP is limited in areas close to magnets; the residual magnetic field has to be lower than 5 mT.

### 3.2.3 *Cryopumping*

Cryopumps rely on three different pumping mechanisms:

- *Cryocondensation.* This mechanism is based on the mutual attraction of similar molecules at low temperature (Fig. 14). The key property is the saturated vapour pressure $P_v$, i.e. the pressure of the gas phase in equilibrium with the condensate at a given temperature [15]. The lowest pressure attainable by cryocondensation pumps is limited by the saturated vapour pressure. Among all gas species, only Ne, $H_2$ and He have $P_v$ higher than $10^{-11}$ Torr at 20 K. The $P_v$ of $H_2$ at the liquid He boiling temperature is in the $10^{-7}$ Torr range, and is $10^{-12}$ Torr at 1.9 K. The quantity of gas that may be cryocondensed is very large and limited only by the thermal conductivity of the condensate.

- *Cryosorption.* This is based on the attraction between gas molecules and substrates (Fig. 14). The interaction forces with the substrate are much stronger than those between similar molecules. As a result, providing the adsorbed quantity is lower than one monolayer, the sojourn time is much longer, and gas molecules are pumped at pressures much lower than the saturated vapour pressure. A significant quantity of gas may be pumped below one monolayer if porous materials are used; for example, in one gram of standard charcoal for cryogenic application, about 1000 m$^2$ of surface are available for adsorption. The important consequence is that significant quantities of $H_2$ and He may be pumped at 20 K and 4.3 K, respectively. In general, submonolayer quantities of all gas species are effectively cryosorbed at their own boiling temperature.

- *Cryotrapping.* In this mechanism, the molecules of a low-boiling-temperature gas are trapped in the condensation layer of another gas. This is possible because the interaction energy between dissimilar molecules may be much higher than that between similar molecules. The trapped gas has a saturated vapour pressure several orders of magnitude lower than in its pure condensate. Typical examples are Ar trapped in $CO_2$ at 77 K and $H_2$ in $N_2$ at 20 K.

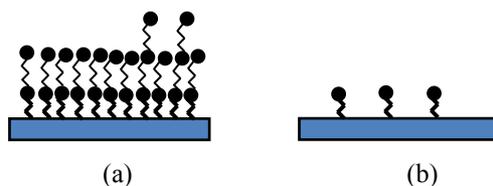

(a)            (b)

**Fig. 14:** Schematic drawing depicting (a) cryocondensation, where the leading mechanism is intermolecular interaction; and (b) submonolayer cryosorption, where the mechanism is molecule–substrate interaction.

Modern cryopumps exploit the first two mechanisms. Cryocondensation takes place on a cold surface, in general at 80 K for water vapour and at 10–20 K for the other gas species. The cryosorption of He, $H_2$, and Ne is localized on a hidden surface coated with a porous material. This part of the pump is kept out of the reach of the other type of molecules, i.e. they have a probability close to 1 of being intercepted and adsorbed by another surface before reaching the cryosorber (see Fig. 15). The refrigeration is obtained by He gas cooled by a Gifford–McMahon cryocooler.

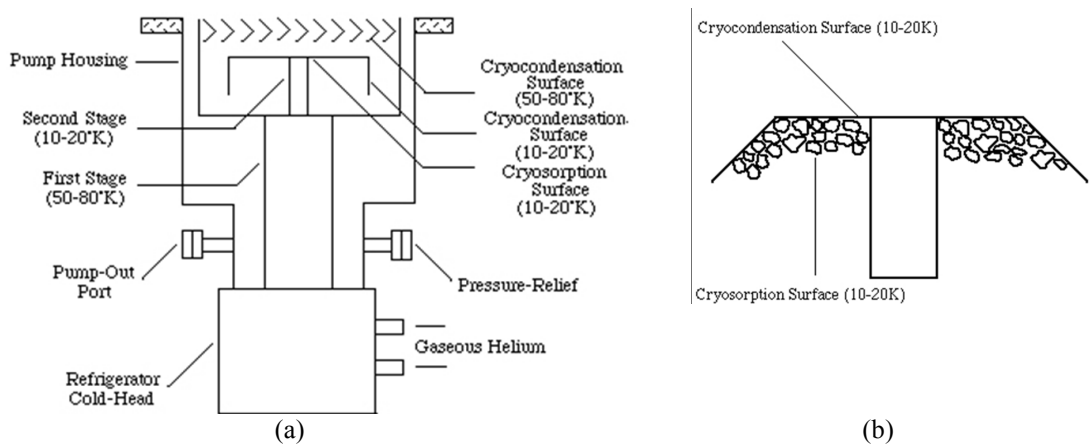

(a)            (b)

**Fig. 15:** Drawings of: (a) a generic cryopump; (b) a closer view to the cryosorption surface where the porous material is fixed.

Cryopumps having pumping speeds in the range 800–60,000 l s$^{-1}$ are commercially available (Table 8). For condensable gas molecules, the capture probability is close to 1 (e.g. for water vapour). The maximum gas capacity (also called maximum gas intake) for the condensable gas is limited only by the thermal conductivity of the condensate. To avoid a thick condensate layer and excessive thermal load, cryopumps should be started in the molecular regime ($P < 10^{-3}$ mbar). The quantity and properties of the porous material determine the maximum gas intake of a cryosorbed gas. In general, it is orders of magnitude lower than that for a condensable gas (see Table 8).

**Table 8:** Pumping speeds and maximum gas capacities of a commercial cryopump (Oerlikon-Leybold 800 BL UHV); the pump inlet diameter is 160 mm.

|                  | H$_2$O | N$_2$   | Ar      | H$_2$ | He  |
|------------------|--------|---------|---------|-------|-----|
| $S$ (l s$^{-1}$) | 2600   | 800     | 640     | 1000  | 300 |
| Capacity (Torr l)|        | 225,000 | 225,000 | 3225  | 375 |

Courtesy of Oerlikon-Leybold.

Cryopumps require periodic regeneration to evacuate the gas adsorbed or condensed; in this way, the initial pumping speed is fully recovered. To do so, the cryopumps are warmed up to room temperature and the released gas is removed by mechanical pumps (mobile TMP in particle accelerators). During regeneration, the pump is separated from the rest of the system by a valve.

Excessive gas adsorption on the cryosorber leads to performance deterioration. A partial and much faster regeneration (1 h against more than 10 h) may be carried out at temperatures lower than 140 K in such a way as to remove the sorbed gas without releasing water vapour from the pump stage at higher temperature.

### 3.2.4 Cryocondensation and cryosorption in the LHC cryogenic vacuum system

In the LHC arcs, pressures of $10^{-8}$ mbar (for hydrogen, measured at room temperature) are needed to fulfil the beam-gas-scattering lifetime requirement. The cold bore, in direct contact with the cold mass of the magnets, is cooled to 1.9 K. At such a temperature, vapour pressures lower than $10^{-12}$ mbar are reached for all gases, except He, for which up to $10^{15}$ molecules cm$^{-2}$ can be cryosorbed before reaching the pressure limit. However, direct interaction between the LHC's proton beams and the cold bore would provoke cryogenic, vacuum, and beam stability issues [16–19].

The adopted solution is the insertion of a tubular shield, called a beam screen (see Fig. 16), between the beam position and the cold bore. The beam screen is cooled by helium gas at a temperature gradually increasing from 5 K to 20 K. The beam screen intercepts the heat flow due to image currents, beam losses, and photon and electron collisions before it reaches the cold mass of the magnets. The drawback of such a solution is the inevitable reduction of the beam pipe aperture.

The beam screen is provided with holes, called pumping slots, to ensure cryocondensation of hydrogen and cryosorption of He on the cold bore at a saturated vapour pressure compatible with the lifetime requirement. The saturated vapour pressures of all other gas species are below the pressure limit between 5–20 K; therefore, they can be efficiently cryocondensated on the beam screen. However, beam-induced photon and electron bombardment on the beam screen can desorb the cryopumped gas, which eventually is cryocondensated on the cold bore after migration through the pumping holes. Owing to the shape and shielding of the pumping slots, the gas pumped on the cold bore is no longer affected by the beam.

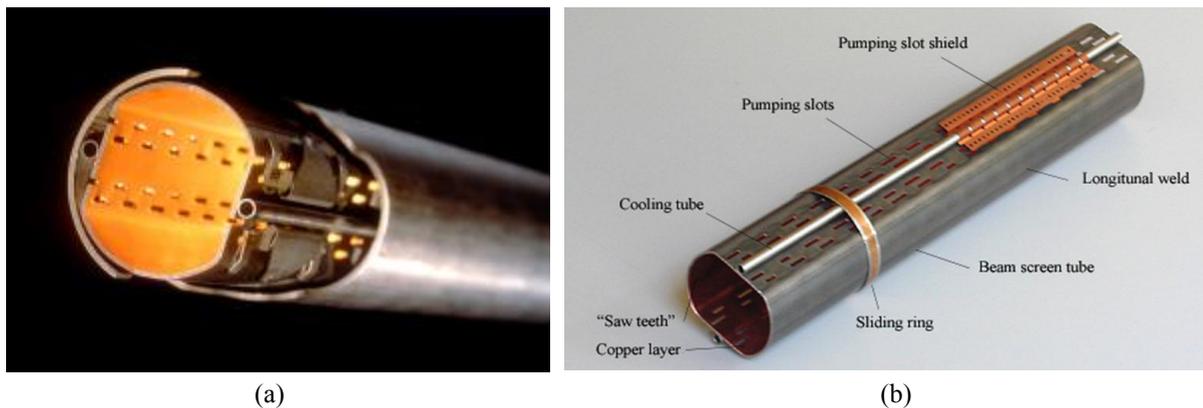

(a)            (b)

**Fig. 16:** Pictures of the LHC's beam screen. (a) The beam screen is shown inserted in a tube having the same diameter of the cold bore. The internal surface is covered by co-laminated copper to reduce the losses generated by image currents. (b) The beam screen outside the cold bore is shown. The pumping slot shield aims at reducing the transmission to the cold bore of electrons and radiation generated by the beam. The two cooling capillaries are laser spot-welded on the beam screen.

In the LHC's long straight sections, the superconducting magnets are cooled at 4.5 K. At this temperature, the saturated vapour pressure of $H_2$ does not fulfil the LHC's pressure requirements. Consequently, $H_2$ has to be pumped by cryosorption onto porous materials as already implemented in cryopumps (see Fig. 15). For such purpose, woven carbon fibre plates developed at the Budker Institute of Nuclear Physics (BINP) are fixed on the external surface of the beam screens [20, 21]. Such a cryosorber may pump more than $10^{17}$ $H_2$ molecules cm$^{-2}$ before reaching equilibrium pressures at $10^{-8}$ mbar, up to 20 K (see Fig. 17). The regeneration, which is possible at temperatures close to 80 K, is done during the magnet warm-ups.

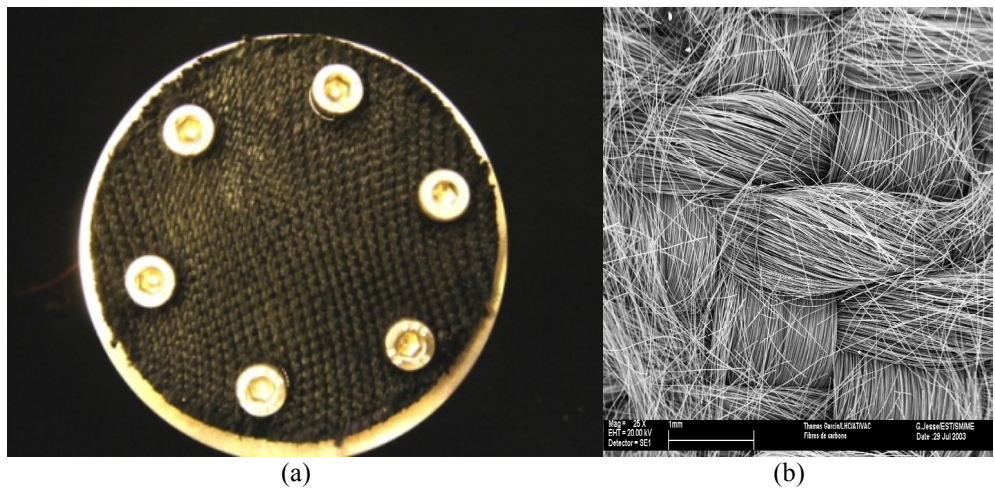

(a)            (b)

**Fig. 17:** Sample of (a) cryosorber made of (b) woven carbon fibres. This material is used to adsorb $H_2$ in the beam pipes of magnets that are cooled at 4.5 K instead of 1.9 K. It is inserted between the pumping slot shields and the beam screen (courtesy of V. Baglin; G. Jesse of CERN MME group made the pictures).